\documentclass[preprint,12pt]{elsarticle}

\usepackage{hyperref,amsmath,amssymb}

\journal{Nuclear Instruments and Methods in Physics Research A}

\begin{document}

\begin{frontmatter}

\title{A Compensated Design of the LGAD Gain Layer}

\author[unito,infn]{V. Sola\corref{corrauth}}
\cortext[corrauth]{Corresponding author}
\ead{valentina.sola@cern.ch}

\author[upo,infn]{R. Arcidiacono}
\author[cnr,infnpg]{P. Asenov}
\author[fbk,tifpa]{G. Borghi}
\author[fbk,tifpa]{M. Boscardin}
\author[infn]{N. Cartiglia}
\author[fbk,tifpa]{M. Centis Vignali}
\author[infnpg]{T. Croci}
\author[upo,infn]{M. Ferrero}
\author[unipg]{A. Fondacci}
\author[unito,infn]{G. Gioachin}
\author[infn]{S. Giordanengo}
\author[unito,infn]{L. Lantieri}
\author[infn]{M. Mandurrino}
\author[unito,infn]{L. Menzio}
\author[unito,infn]{V. Monaco}
\author[infnpg]{A. Morozzi}
\author[cnr,infnpg]{F. Moscatelli}
\author[unipg,infnpg]{D. Passeri}
\author[infn]{N. Pastrone}
\author[fbk,tifpa]{G. Paternoster}
\author[unito,infn]{F. Siviero}
\author[infn]{A. Staiano}
\author[unito,infn]{M. Tornago}

\address[unito]{Universit\`a degli Studi di Torino, via P. Giuria 1, 10125, Torino, Italy}
\address[infn]{INFN, Sezione di Torino, via P. Giuria 1, 10125, Torino, Italy}
\address[upo]{Universit\`a del Piemonte Orientale, largo Donegani 2/3, 28100, Novara, Italy}
\address[cnr]{CNR-IOM, Sede di Perugia, via A. Pascoli, 06123, Perugia, Italy}
\address[infnpg]{INFN, Sezione di Perugia, A. Pascoli, 06123, Perugia, Italy}
\address[fbk]{Fondazione Bruno Kessler, via Sommarive 18, 38123, Povo (TN), Italy}
\address[tifpa]{TIFPA-INFN, via Sommarive 18, 38123, Povo (TN), Italy}
\address[unipg]{Universit\`a degli Studi di Perugia, A. Pascoli, 06123, Perugia, Italy}

\begin{abstract}

  In this contribution, we present an innovative design of the Low-Gain Avalanche Diode (LGAD) gain layer,
  the {\it p$^+$} implant responsible for
  the local and controlled signal multiplication. In the standard LGAD design, the gain layer is obtained by implanting
  $\sim$ 5E16/cm$^3$ atoms of an acceptor material, typically Boron or Gallium, in the region below the {\it n$^{++}$} electrode.
  In our design, we aim at designing a gain layer
  resulting from the overlap of a {\it p$^+$} and an {\it n$^+$} implants: the difference
  between acceptor and donor doping will result in an effective concentration of about 5E16/cm$^3$, similar to standard LGADs.
  At present, the gain mechanism of LGAD sensors under irradiation is maintained up to a fluence of $\sim$ 1--2E15/cm$^2$, and then it is
  lost due to the
  acceptor removal mechanism.~The new design will be more resilient to radiation, as both acceptor and donor atoms will undergo removal
  with irradiation, but their difference will maintain constant.~The
  compensated design will empower the 4D tracking ability typical of the LGAD sensors well above 1E16/cm$^2$.
  
\end{abstract}

\begin{keyword}
 Silicon Sensor \sep  LGAD \sep Compensation \sep Compensated LGAD \sep Gain Layer \sep 4D Tracking \sep Radiation Hardness
\end{keyword}

\end{frontmatter}



\section{Introduction}
\label{sec:intro}

Thin Low-Gain Avalanche Diodes (LGADs)~\cite{Pell1}, with an active thickness of $\sim$ 50 $\mu$m, have demonstrated excellent
performances in timing measurements~\cite{UFSD3}. Therefore, LGAD sensors are good candidates to perform
4D tracking~\cite{UFSD2}, due to their ability to combine precise timing with precise tracking measurements.

\begin{figure}[b!]
\centering
\includegraphics[width=1\linewidth]{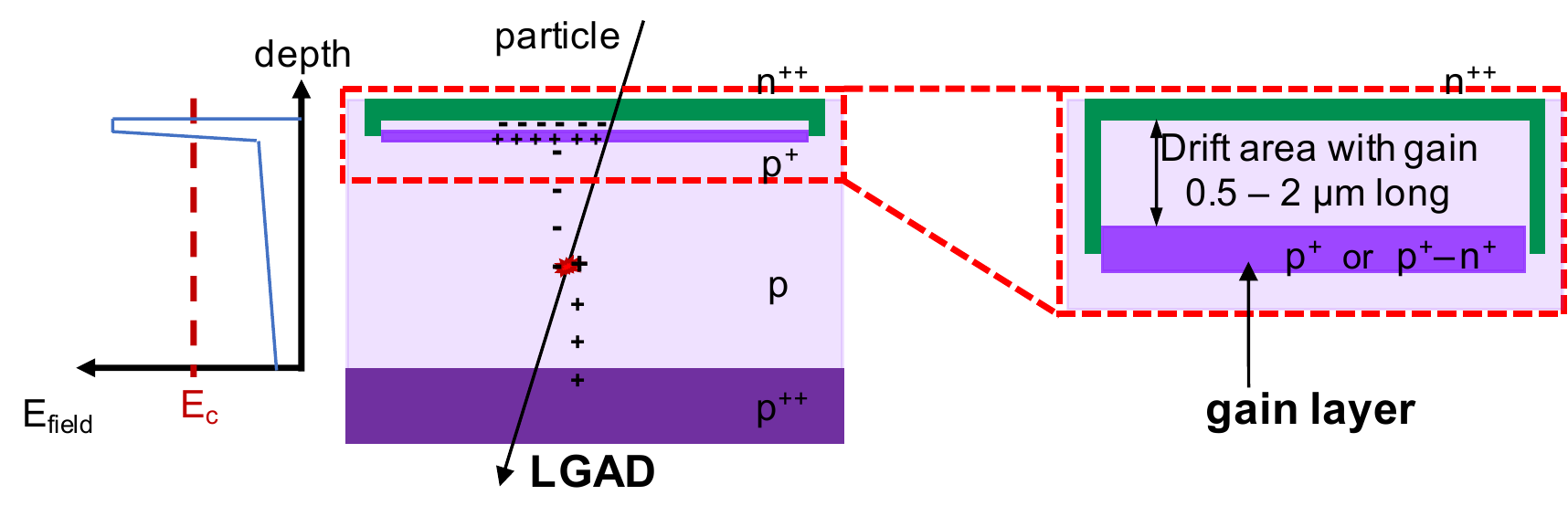}
\caption{Sketch of an LGAD sensor. Impact ionisation occurs for E$_{field} \gtrsim $ E$_c$. }
\label{fig:LGAD}
\end{figure}

Figure~\ref{fig:LGAD} shows a sketch of an LGAD sensor:
the high resistivity {\it p} bulk readout through an {\it n$^{++}$} implant ({\it n-in-p} sensor)
ensures good stability and control of the electric field,
while the thin {\it p$^+$} layer, the so-called gain layer, with a high acceptor concentration of the order of 5E16
atoms/cm$^3$ provides a local increase of the electric field, where controlled signal multiplication occurs.
However, due to acceptor removal, the \textit{p$^+$} doping concentration of the gain implant gets reduced by irradiation and at a
fluence of 1--2E15/cm$^2$ the LGADs lose their multiplication power and behave as standard \textit{n-in-p} sensors \cite{UFSD2irr}.
 
\begin{figure}[t!]
\centering
\includegraphics[width=1\linewidth]{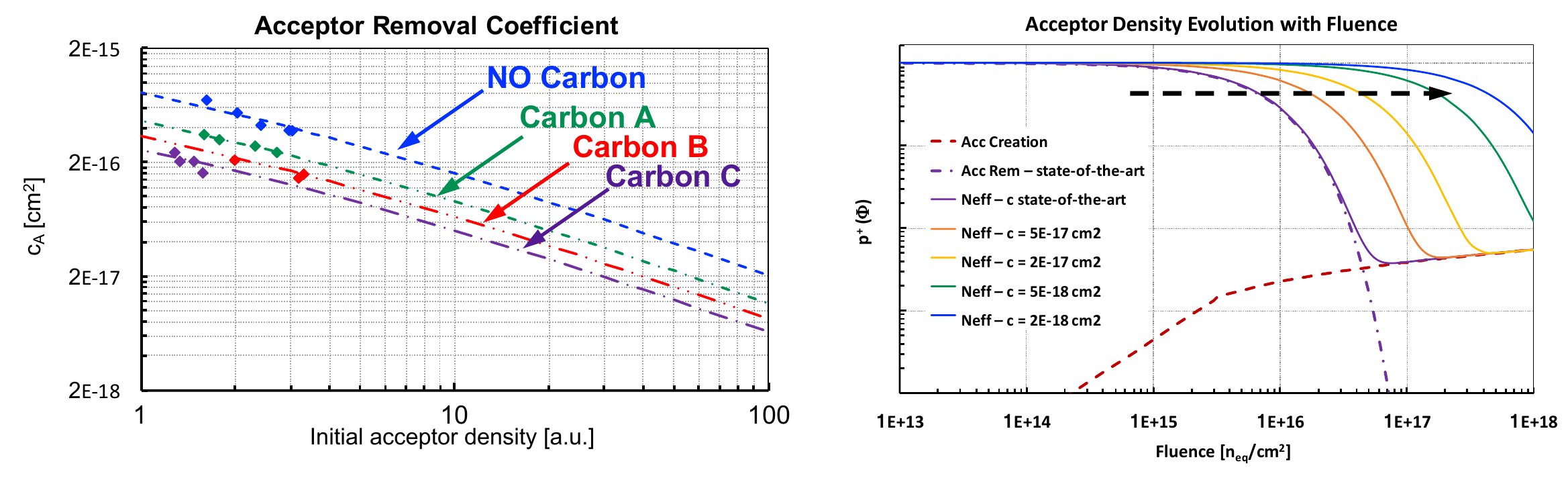}
\caption{Evolution of the acceptor removal rate coefficient for different strategies of Carbon co-implantation~\cite{UFSD2irr} (left),
  and acceptor survival probability with fluence (right).}
\label{fig:cevo}
\end{figure}

Acceptor removal is the transformation of electrically active dopant atoms into neutral defect complexes. The
removal by irradiation has been measured using different initial acceptor densities, N$_{A}$(0), and the effective
doping concentration is exponentially dependent on the irradiation fluence, $\Phi$, according to:
\begin{equation}
  N_A(\Phi) = N_A(0) \cdot e^{- c_A \cdot \Phi } ~,  
\label{eq:removal}
\end{equation}
where $c_A$ is the acceptor removal coefficient, which depends on the initial concentration:
a higher initial concentration  N$_{A}$(0) results in slower acceptor removal c$_{A}$,
hence the probability to experience
removal under irradiation decreases.
The addition of specific defects can mitigate the removal:
the addition of carbon atoms slows the removal of acceptors, and different concentrations and
activation strategies can increase radiation tolerance by a factor of 3, as shown in Fig.~\ref{fig:cevo} (left).
 
A new paradigm for the gain layer design is necessary to overcome the present limit of radiation tolerance for the gain implant,
to preserve internal signal multiplication up to the highest fluences. Figure~\ref{fig:cevo} (right) shows
that in orded to extend the multiplication power up to
1E17/cm$^2$ and beyond, it is necessary to reduce the acceptor removal coefficient down to the values of $\sim$ 5E--18 cm$^2$:
with the present gain layer design, the doping of the {\it p$^+$} implant has to be increased by more than a factor of 10 to reach the target value of
c$_A$. This is an unfeasible approach since the gain generated by such a high-density implant will be too high to be controlled by the external bias.

\section{The compensated LGAD design}
\label{sec:comp}

\begin{figure}[t!]
\centering
\includegraphics[width=1\linewidth]{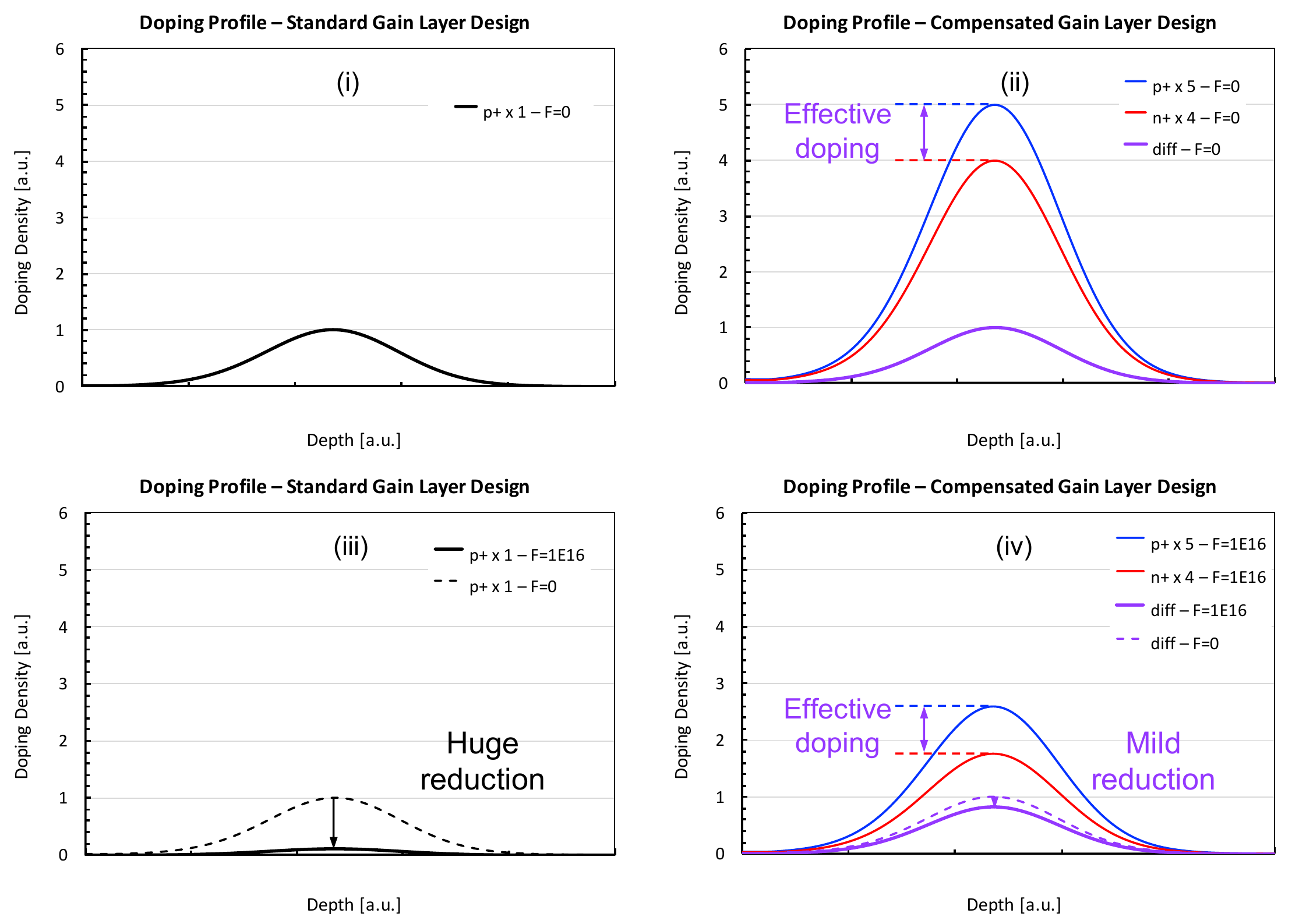}
\caption{Evolution with irradiation of a Gaussian gain layer profile in standard (left) and compensated (right) LGAD.
  The absolute and effective doping densities are shown before irradiation (top) and after a fluence of 1E16/cm$^2$ (bottom).}
\label{fig:GLevo}
\end{figure}

To overcome the current limitations of the standard gain layer design,
we introduce an innovative design of the LGAD gain layer,
the {\it compensated gain layer}.
This breakthrough is described graphically in Fig.~\ref{fig:GLevo}.
As explained in Sect.~\ref{sec:intro}, in a standard LGAD, the doping density of the gain implant (\textit{p}-doped)
decreases with irradiation due to the process of acceptor removal, Fig.~\ref{fig:GLevo} (i) and (iii). In
the new design, which we call compensated LGAD,
the effective doping of the gain implant is obtained by combining a \textit{p$^+$}-doped and an \textit{n$^+$}-doped implant. If
implemented correctly,
the \textit{n$^+$} doping compensates part of the \textit{p$^+$} doping,
leaving an effective \textit{p$^+$--n$^+$} doping similar to that used in the standard
design, Fig.~\ref{fig:GLevo} (ii).
Both \textit{p$^+$} and \textit{n$^+$} implants will experience removal by irradiation,
but, if properly engineered, their difference can maintain nearly constant with irradiation,
as it is sketched in Fig.~\ref{fig:GLevo} (iv).
 
Donor atoms experience removal under irradiation following the same law as acceptors, see Eq.~(\ref{eq:removal}).
However, the donor removal coefficient, $c_D$, is unknown at the
donor densities of interest for the compensated LGAD design~\cite{DonorRem}.
Therefore, under irradiation, three scenarios are possible:
\begin{itemize}
\item[(i)] The acceptor and donor removal coefficients are of the same order, $c_A \sim c_D$, and their difference will remain constant,
  yielding an unchanged gain with irradiation. This scenario represents the best possible outcome.
\item[(ii)] The acceptor-atoms removal is faster, with $c_A > c_D$. Albeit not as good as (i), this option still leads
  to higher radiation resistance, as the overall rate of effective doping disappearance is slower than in the standard design. As
  an additional handle to be used in this case, co-implantation of carbon atoms can be used to mitigate the removal of the
  {\it p$^+$}-doping~\cite{UFSD2irr}. So, the addition of carbon can turn scenario (ii) into (i).
\item[(iii)] The donor-atoms removal is faster, $c_A < c_D$, resulting in a rapid increase of the net {\it p$^+$}-doping. In
  this case, for equal bias voltage,
  the gain increases with irradiation. This situation can be handled by a decrease of the operational bias voltage or by adding oxygen
  to the {\it n$^+$} implant,
  to decrease the donor deactivation rate~\cite{MMoll1}.
\end{itemize}

It will be necessary to investigate the donor removal at a presently unexplored region of donor densities,
between $\sim$ 1E16 and 1E17 atoms/cm$^3$, as in literature $c_D$ studies are limited to densities up to 1E14 atoms/cm$^3$
\cite{DonorRem,MMoll1}.
Moreover, a dedicated study of the optimal parameters of oxygen co-implantation, such as density and temperature of activation,
is needed to allow a level of understanding similar to the one achieved for carbon co-implantation, shown in Fig.~\ref{fig:cevo} (left).
A further unknown which will require inspection is the interplay between acceptor and donor removal mechanisms when high densities
of {\it p} and  {\it n} atoms are implanted in the same volume.

\section{Simulation of the compensated gain layer}
\label{sec:sim}

\begin{figure}[b!]
\centering
\includegraphics[width=.5\linewidth]{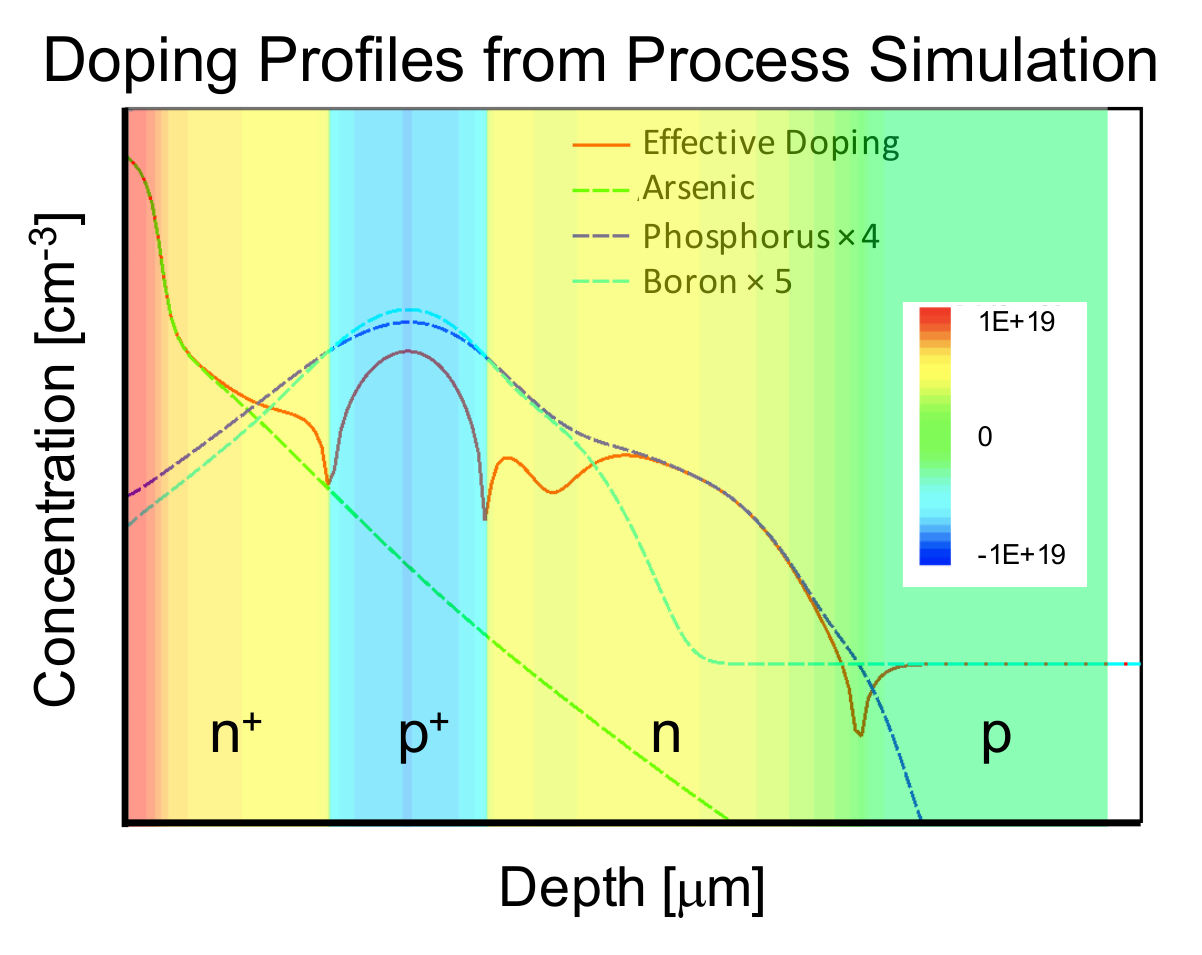}
\caption{Process simulation of the gain layer implants for a compesated LGAD. The {\it n$^{++}$} contact is included in the simulation.}
\label{fig:process}
\end{figure}

A simulation of the production process of compensated LGAD sensors has been performed using TCAD Silvaco~\cite{silvaco}.
Boron has been chosen as {\it p$^+$} dopant, and Phosphorus as the {\it n$^+$} counterpart. The {\it n$^{++}$} contact
is made of Arsenic. Different implantation energies and 
activation temperatures have been tested. One possible outcome of the process simulation is shown in Fig.~\ref{fig:process}, where
4 parts of Phosphorus are used to balance 5 parts of Boron: due to the higher atomic
weight of Phosphorus atoms, the {\it n$^+$} profile exhibits a deeper tail with respect to the {\it p$^+$} one.

\begin{figure}[t!]
\centering
\includegraphics[width=0.5\linewidth]{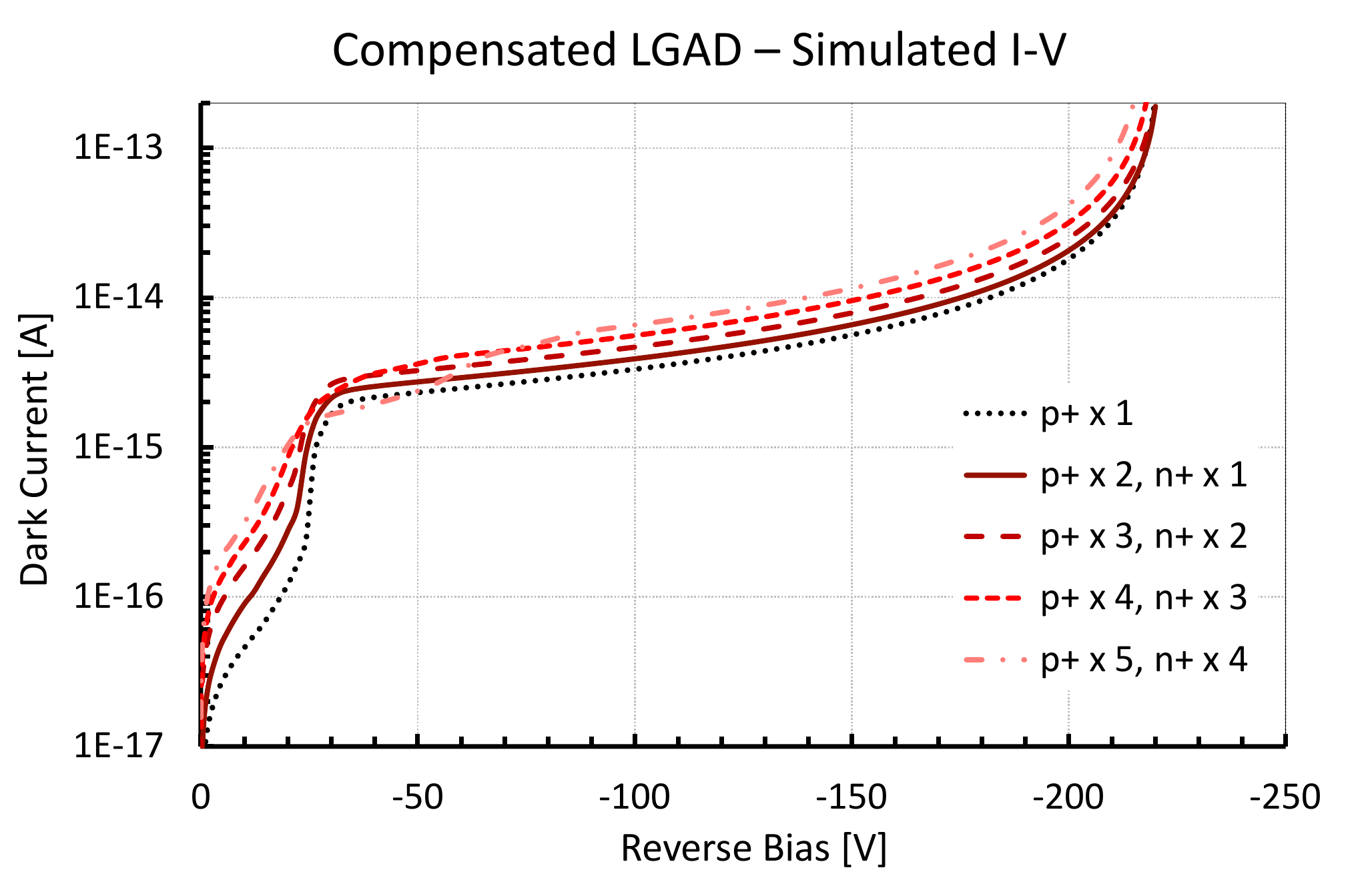}
\caption{Simulated current-voltage characteristics for compensated LGAD sensors with different combinations of
  {\it p$^+$--n$^+$} doping concentration. The current-voltage characteristic of a standard LGAD sensor is
  included for comparison.}
\label{fig:IVsim}
\end{figure}

The steady state behaviour of the compensated LGAD sensors has been simulated
using Sentaurus Synopsys TCAD~\cite{synopsys}, to investigate
the overall performances of the new design and how the
doping transition regions observed in the gain layer volume (Fig.~\ref{fig:process})
can affect the sensors' operation.
In Fig.~\ref{fig:IVsim},
the dark current evolution as a function of the reverse bias has been simulated for different combinations of 
{\it p$^+$--n$^+$} dopant compensation: with proper tuning of the acceptor and donor concentrations
it is possible to adjust the voltage at which the breakdown is reached.
The simulations demonstrated that even using different
concentrations of Boron and Phosphorus, the compensated LGADs behave similarly to standard LGAD, and with a proper 
design of the {\it p$^+$} and {\it n$^+$} implants, it is possible to 
precisely tune the current-voltage characteristic and the break-down voltage of the sensors.

\section{Conclusion}
\label{sec:sum}

A new paradigm for the design of the gain layer has been proposed, based on the compensation of
{\it p$^+$} and {\it n$^+$} dopants, which results in an effective {\it p$^+$} doping profile similar to the standard
LGAD design.

With appropriate optimisation of the process parameters,
this innovative design will allow the usage of 
LGAD sensors for 4D tracking 
well above the present limit in fluence of about
1--2E15/cm$^2$: both the {\it p$^+$} and {\it n$^+$} implants will suffer from dopant removal, however, their
difference will keep almost constant under irradiation, preserving the multiplication power of the LGAD sensors up to the
highest fluences.

The first batch of compensated LGAD sensors is currently in production at the
Fondazione Bruno Kessler (FBK)foundry~\cite{fbk}, and
its completion is expected by autumn 2022: it
will assess the feasibility of the design and allow for the identification of the best parameters
to manufacture compensated LGAD sensors.

\section*{Aknowledgments}

This project has received funding from
the INFN CSN5 through the `eXFlu' research project,
the PRIN MIUR project 2017 2017L2XKTJ `4DInSiDe',
and the European Union’s Horizon 2020 Research and Innovation programme under Grant Agreement No. 101004761 (AIDAinnova).

\section*{References}

\bibliography{vs_bibfile}

\end{document}